\documentstyle[aps,epsfig]{revtex}
\begin{document}
\title{Heavy flavor production at RHIC and LHC energy}

\author{\bf A. K. Chaudhuri\cite{byline}}
\address{ Variable Energy Cyclotron Centre\\
1/AF,Bidhan Nagar, Kolkata - 700 064\\}
\maketitle
\begin{abstract}

In a leading order pQCD model, we have studied the heavy flavor
production in p+p collisions 
at RHIC and LHC energy. Leading order pQCD models require
a K-factor. At RHIC energy, $\sqrt{s}$=200 GeV, we fix K such that the model reproduces the integrated charm yield, $dN^{c\bar{c}}/dy$, estimated by the STAR and the PHENIX collaboration 
in p+p collisions.
The model then explains the  STAR   data
on the
transverse momentum distribution of open charm mesons $(D^0)$
and decay electrons in p+p collisions.
The p+p predictions, scaled by the number of binary
collisions, also explain the electron spectra in STAR p+d collisions
and PHENIX Au+Au collisions 
in different centrality bins.
Assuming that at LHC energy K-factor is of the order of unity, we have
used the model to predict the transverse momentum distribution
of $D$ and $B$ mesons and also of electrons from semileptonic decay of $D\rightarrow e$
and $B\rightarrow e$, in p+p collisions at LHC energy, $\sqrt{s}$=14 TeV.
 
\end{abstract}


\section{introduction}

Recently STAR collaboration has measured the transverse momentum
distribution of open charm mesons, $D^0(\bar{D}^0)$ from
direct reconstruction of $D^0 (\bar{D}^0) \rightarrow K^\mp \pi^\pm$
in d+Au collisions and indirect electron-position measurements via
charm semileptonic decays in p+p and d+Au collisions at 
$\sqrt{s}$=200 GeV \cite{star1}. Electron spectra show approximate
binary collision scaling between p+p and d+Au collisions. The total
charm cross section per nucleon-nucleon interaction for d+Au collisions
at $\sqrt{s}$=200 GeV is $\sigma^{c\bar{c}}=1.3 \pm 0.2 \pm 0.4$ mb,
considerably larger than the standard PYTHIA prediction.
Semileptonic decay spectra of electrons has also
been measured by the PHENIX collaboration \cite{phenix1} in p+p and
in Au+Au collisions, at $\sqrt{s}$=200 GeV. 
In Au+Au collisions, PHENIX measured charm electrons in different centrality
bins.
PHENIX data also indicate that the binary collision scaling holds for charm
production in Au+Au collisions. However, PHENIX
measured lesser number of charm. Their estimate,
$\sigma^{c\bar{c}}=622 \pm 57 \pm 160$ $\mu$b, is approximately 
one half
of the STAR estimate.
PHENIX Au+Au charm results are in contrast with the strong violation of binary collision scaling
observed in non-charmed (light) hadrons.
All the four RHIC experiments, STAR, PHENIX, PHOBOS and
BRAHMS reported large suppression of high $p_T$ (light) hadrons in Au+Au collisions
\cite{star2,phenix2,phobos1,brahms1}. High $p_T$
suppression of non-charmed (Light) hadrons are generally explained in terms of
partonic energy loss in a dense medium \cite{gy90,wa92}. Indeed, they are the most
important inputs for the claim that very high density medium is created at RHIC Au+Au collisions.
However the charm results are not contradictory with the high $p_T$ suppression observed in
non-charmed hadrons.
Energy loss do not reduce the number
of partons, rather change the $p_T$ distribution.  
Thus if in Au+Au collisions, heavy quarks suffered energy loss,  momentum distribution of
charmed electrons would not have scaled with charmed electron distribution in pp collisions. The PHENIX
Au+Au data on charm electrons
then confirm the believe that
heavy quarks suffer no or little energy loss in the medium. 

Theoretical understanding of heavy flavor production  is of considerable interest.
Perturbative QCD is  better
applicable for heavy flavor production.  Their large mass provide a 
natural scale for perturbative expansion.
Experimental data on heavy flavor
production can test the perturbative models and provide important
inputs like various mixing angles. In heavy ion collisions at RHIC,
heavy flavor production can give important information  about the
initial condition of the medium produced. 
Due to their large mass, heavy flavors are 
produced in initial hard collisions and in a short time scale and are
ideal probe for the initial condition of the medium produced.
Passage of a heavy flavor through
a deconfined medium can alter the momentum distribution of heavy 
flavored hadrons. Semileptonic decay of a heavy flavored hadron 
$D(B)\rightarrow Xe\nu$, add to the 
charmonium background, an important
signal of the confinement-deconfinement phase transition. 
Open charm yield is also important for understanding charmonium 
production.
Quite early, Matsui and Satz \cite{ma86} suggested that charmoniums will be 
suppressed in a deconfined medium.  However, recently
there have been suggestions that rather than suppression, 
at RHIC, charmonium yield will be enhanced due to recombination effect
\cite{th01}. Open charm yield can help to decipher the issue of
suppression/enhancement of charmoniums.
 
Heavy flavor production in pQCD models
has been studied earlier also \cite{ca98,vo02,ca05}. 
Recently
Cacciari et al \cite{ca05} made an up-to-date prediction for open charm and bottom
production at RHIC p+p collisions, using the First Order plus Next-to-leading-Log (FONLL) level. 
In the present paper, we have studied heavy flavor production 
in  a leading order pQCD model. 
Leading order pQCD models
are simpler but need a $K$-factor. $K$ factor accounts for the
neglect of all the higher order terms in the perturbative expansion. 
K-factor depends on energy. Energy dependence of $K$-factor has
been studied in \cite{es03}.
$K$-factor
decreases with energy and at RHIC energy, $K \sim 3.4$.  At still higher
energy $K$ factor approaches unity.
In the present paper, at RHIC energy K is obtained by fitting the integrated charm
yield $dN^{c\bar{c}}/dy$ in STAR and PHENIX p+p collisions. With the K-factor
so fixed, the model is used to explain the STAR and PHENIX data on the 
transverse momentum distribution of
$D$-mesons and decay electrons. As it will be shown later, the model, 
with heavy quark fragmentation function, parameterised by Braaten et al \cite{br95},
give excellent description to STAR and PHENIX data.
The more well known Peterson's fragmentation function \cite{pe83} 
underpredict the data by a factor of 3 or so.
The model can serve as a baseline for comparing heavy flavor data in
RHIC Au+Au collisions and in future LHC energy collisions. 

The paper is organised as follows: in section 2, we briefly describe the
pQCD model. Results for heavy flavor production and comparison
with RHIC experiments (STAR and PHENIX) are done in section 3. Model predictions
for heavy meson  production and decay electrons at LHC energy, $\sqrt{s}$=14 TeV p+p collisions are also given in sec.3.
Summary and conclusions are given in section 4.
 
\section{pCQD model of charm quark and charmed meson production}

In pQCD models, semileptonic decay of a heavy meson, occur in three step:
(i) production of  a heavy quark in parton level collisions, (ii) fragmentation
of the heavy quark in to  a heavy mesons and (iii) semileptonic decay of
heavy meson. Schematically 
transverse momentum spectra of decay electrons in p+p collisions
can be written as (schematically),

\begin{equation} \label{1}
\frac{d\sigma^e}{dydp^2_T}=\frac{d\sigma^Q}{dy_Qd^2p_{T_{Q}}}
\bigotimes D(Q\rightarrow H_Q) \bigotimes f(H_Q\rightarrow e)
\end{equation}

\noindent where the symbol $\bigotimes$ denote generic convolution
and
$D(Q\rightarrow H_Q)$ and $f(H_Q \rightarrow e)$
represent the fragmentation of
heavy quark Q into a heavy flavored hadron $H_Q$, and semileptonic decay of $H_Q$ into electron respectively.

Detail of the evaluation of Eq.\ref{1} is given in the appendix. 
For evaluation of Eq.\ref{1}, parton distribution functions and fragmentation
function of a heavy quark to fragment into a heavy hadron, are required.
For the parton distributions, we have used CTEQ5L parameterisation.
Several fragmentation function
have been used in literature \cite{br95,pe83,ka78,co85}. Most well known is the parameterisation by Peterson et
al \cite{pe83}.
Peterson's parameterisation does not contain spin information and is parameterised as,

\begin{equation} \label{2}
D_{c/C}(z)=\frac{N}{z[1-1/z-\varepsilon_Q/(1-z)]^2}
\end{equation}

\noindent here $N$ is the normalisation and $\varepsilon_Q$ is a parameter.
The parameter $\varepsilon_Q$ is approximately the ratio
of square of constituent light quark (q) and heavy quark (Q) masses, $\varepsilon_Q=m^2_q/m^2_Q$. $\varepsilon_Q=m^2_q/m^2_Q$
depend on the heavy quark mass.  
Most commonly
used values are,
$\varepsilon_Q=0.5$ for fragmentation of a charm quark into a D-meson and
$\varepsilon_Q=0.006$ for fragmentation of a bottom quark into a B-meson.
More recently Braaten et al \cite{br95} parameterized the heavy quark
fragmentation function. It contains the spin information.
The pseudo-scalar and vector meson fragmentation functions are 
written as,

\begin{eqnarray} \label{3}
D_{Q\rightarrow P}(z)=&&N \frac{rz(1-z)^2}{[1-(1-r)z]^6}
[6-18(1-2r)z+(21-74r+68r^2)z^2-2(1-r)(6-19r+18r^2)z^3+ \nonumber \\ 
&&3(1-r)^2(1-2r+2r^2)z^4]\\ \nonumber
D_{Q\rightarrow V}(z)=&&N \frac{rz(1-z)^2}{[1-(1-r)z]^6}
[2-2(3-2r)z+3(3-2r+4r^2)z^2-2(1-r)(4-r+2r^2)z^3+\\
&&(1-r)^2(3-2r+2r^2)z^4]
\end{eqnarray}

Parameter $r$ is approximately the ratio of constituent quarks masses, $r=m_q/m_Q$. $r=0.1$ for c-quark
fragmentation to D-mesons, and $r=0.03$  for fragmentation of $b$-quarks into a B-meson.
In the present work we have used the pQCD pseudo-scalar fragmentation functions parameterised by
Braaten et al \cite{br95}, henceforth call FF-I.
For comparison, 
we will present results obtained with the fragmentation function
parameterised by
Peterson et al \cite{pe83} (henceforth call FF-II). As will be shown later,
fragmentation function FF-II yields are nearly a factor of two less than
the yield obtained from FF-I. Peterson's fragmentation function is
comparatively hard.

Fragmentation functions, Eq.\ref{2} and \ref{3} are used as the initial distribution at momentum scale,
$q^2=m^2_{c(b)}$. In the present work, we have used $m_c$=1.5 GeV
and $m_b$=4.4 GeV as charm and bottom quark mass respectively.
Fragmentation function at other momentum scales are obtained from
numerically integrating the Alterelli- Parisi evolution equation,
 
\begin{equation}
\mu^2 \frac{\partial}{\partial \mu^2} D_{Q_H}(z,\mu^2) =
\int_z^1 \frac{dy}{y} P_{Q\rightarrow Q}(\frac{z}{y},\mu) D_{Q\rightarrow H}(y,\mu^2)
\end{equation}

\noindent where $P_{Q\rightarrow Q}$ is the splitting function:

\begin{equation}
P_{Q\rightarrow Q}(z,\mu^2)=\frac{2\alpha_s(\mu^2)}{3\pi}
\left(\frac{1+z^2}{1-z)} \right)_+
\end{equation}
 
\begin{figure}[h]
\begin{minipage}{18pc}
\includegraphics[bb=30 250 520 800,width=15pc,height=15pc]{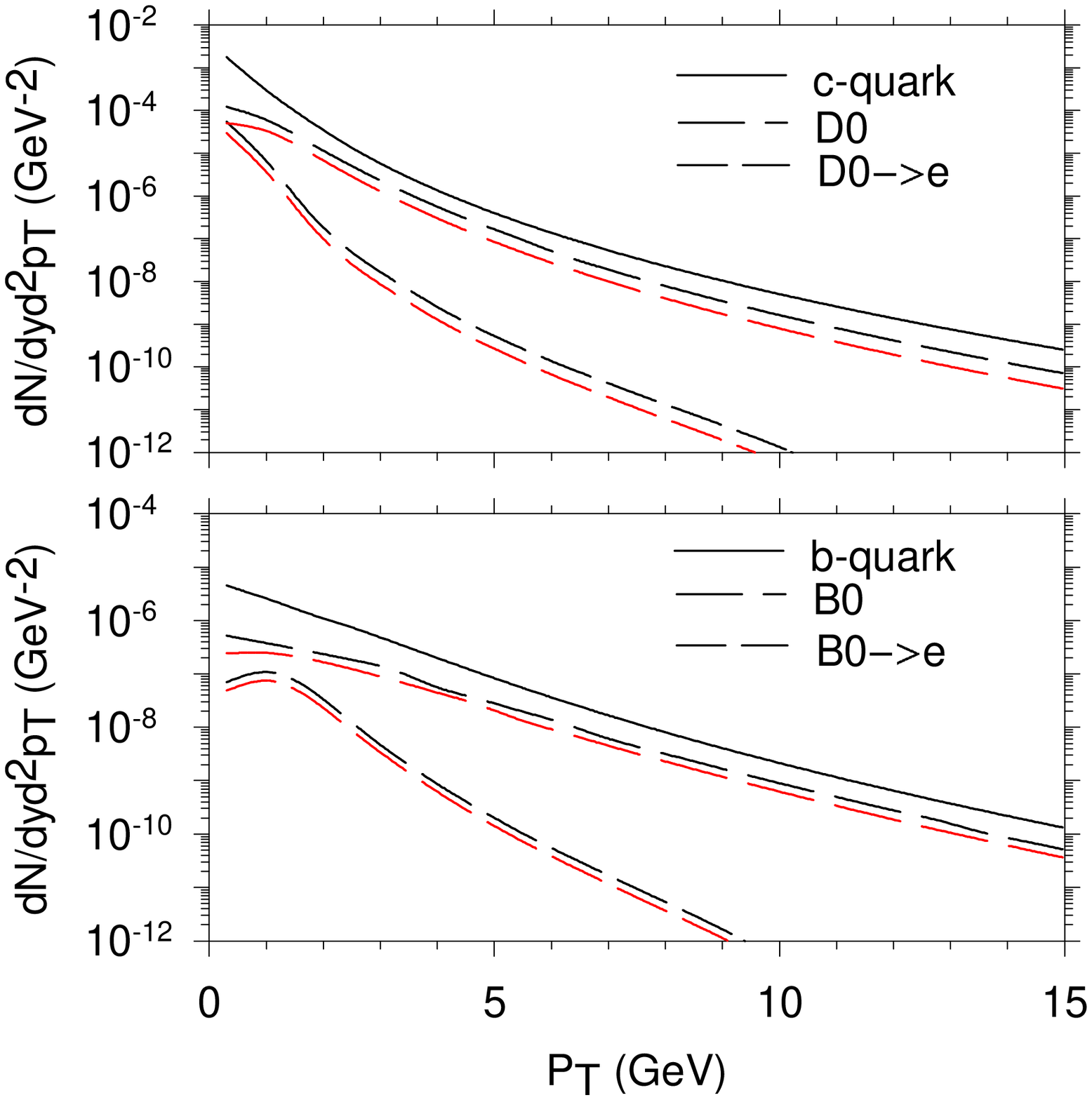}
\caption{\label{F6} 
(a) Solid line is the pQCD prediction for the $p_T$ distribution of $c$-quarks at
$\sqrt{s}$=200 GeV with K=1.
The dashed and short-dashed lines are the model predictions for $D^0$-mesons and decay electrons with
FF-I. Red dashed line and short
dashed lines are the model predictions for D-meson and decay electrons
with FF-II. (b) same for b-quarks, B-mesons and $B\rightarrow e$ decay.}
\vspace{.5cm}
\end{minipage}\hspace{2pc}%
\begin{minipage}{18pc}
\includegraphics[bb=30 250 520 800,width=15pc,height=15pc]{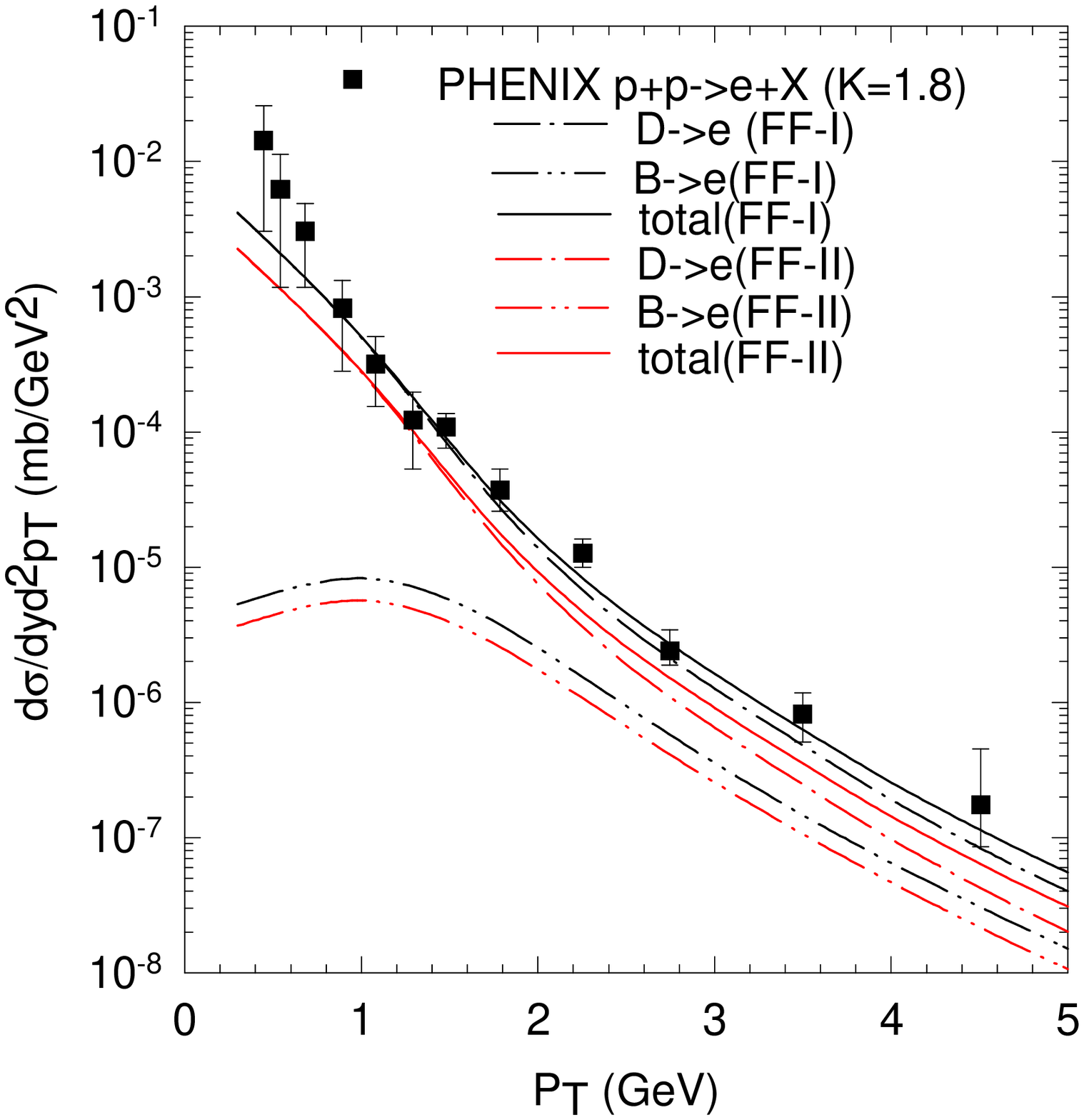}
\caption{\label{F7} PHENIX measurements of 
$p_T$ distribution of charm electrons in p+p collisions at RHIC energy, 
$\sqrt{s}$=200 GeV. The pQCD predictions
with FF-I , are shown as
dash-dot and  dash-dot-dot lines, for D and B meson decay respectively.
  The solid line is the sum of the two contributions.
Results obtained with FF-II are shown as red lines. The K-factor is 1.8.}
\vspace{8mm}
\end{minipage} 
\end{figure}

\section{results}

\subsection{Transverse momentum distribution of $D^0$,$B^0$ and 
semileptonic decay electrons at RHIC} 

The pQCD model predictions for the transverse momentum distribution of
heavy mesons and decay electrons in $\sqrt{s}$=200 GeV p+p collisions are shown in Fig.1a and
1b. Here we have taken  K=1. In Fig.1a, solid black line is the $p_T$
spectrum for charm quarks. Charm quark fragmentation to D mesons and
semileptonic decay of D-mesons, obtained with the fragmentation function FF-I is shown as black dashed and short-dashed lines respectively.
The same obtained with the fragmentation function FF-II are shown as red dashed and red short-dashed lines.
$p_T$ distribution of $D$-mesons closely scale with c-quarks but
$p_T$ distribution of electrons falls much faster than that of c-quarks
or of $D$-mesons. We also note that yield of $D$ mesons and also of
decay  electrons, obtained with the fragmentation function FF-I  are 
approximately a factor of 2 larger than the
yield obtained with the parameterisation FF-II.
Peterson fragmentation function (FF-II) is comparatively hard.
Same results, for bottom production are shown in Fig.1b. 
At RHIC energy,
low $p_T$ heavy quarks are dominantly charm. Heavy mass of bottom
quarks inhibit their production at low $p_T$.
However, at large $p_T$, bottom
production evenly competes with charm production. Consequently, at
large $p_T$ semileptonic decay electrons will have contributions from
$D$ as well as from $B$ mesons.
For B-mesons also fragmentation function FF-II 
produces factor of two less number of B-mesons than the fragmentation function FF-I.

As mentioned earlier,    leading order 
pQCD models require a k-factor. STAR and PHENIX collaboration had estimated
the total charm yield in p+p collisions at $\sqrt{s}$=200 GeV. While the
PHENIX collaboration \cite{phenix1} obtained $d\sigma^{c\bar{c}}/dy=143 \pm 13 \pm 36$ $\mu$b in the $p_T$
range,
$0.4 GeV <p_T<4 GeV$, the STAR collaboration \cite{star1} obtained larger value,
$d\sigma^{c\bar{c}}/dy=0.29 \pm 0.04 \pm 0.08$ mb in $p_T$ range,
$1 GeV <p_T<4 GeV$. 
We have fixed the K factor such that the integrated charm yield in the model reproduces the central value of the
PHENIX and STAR estimate. We obtain, K=1.8 and K=7.0 for PHENIX 
and STAR experiment respectively.
We could have fixed K to reproduce the $D$ meson or the electron yield.
But fixing the K from charm yield has the advantage that the model can
now test different fragmentation functions by comparing the model
predictions with experiment.
Before we continue we note that the large
uncertainty in K (K=1.8 for PHENIX experimental data and 7.0 for STAR experimental data)is a direct result of large difference in integrated charm cross sections in two measurements. We hope that in future the large
difference in charm cross sections, in two experiments, will be removed.
and K can be fixed more accurately.

With the only parameter of the model fixed, we can compare model 
predictions with other experimental observable, namely, $p_T$ distribution
of $D^0$ and decay electrons in STAR and PHENIX experiments.
In Fig.2, $p_T$ distribution of charm electrons in pp collisions, measured
by the PHENIX collaboration \cite{phenix2} at $\sqrt{s}$=200 GeV  
is shown. Present pQCD model predictions obtained with the
fragmentation
functions FF-I (black lines) and FF-II (red lines)
are also shown in Fig.2. For both the fragmentation functions, the 
K-factor is fixed, K=1.8. We have shown 
the contribution of both D (the dash-dot line) and B mesons (dash-dot-dot line) decay. Solid lines are the sum of the two contributions.
At low $p_T$, decay of $B$-mesons contribute negligibly to the total
electron spectrum. For $p_T <$2 GeV, bottom contribute 
less than 10\% to the total yield. In the $p_T$ range 2-3 GeV bottom
contribute 10-20\% and in $p_T$ range 3-4 GeV bottom contribution rises
to 20-30\%. This numbers compare favorably with PHYTHIA estimate
\cite{sj01}.  
Present pQCD model with FF-I well reproduces the PHENIX data. 
On the other hand fragmentation function FF-II, though
reproduce the shape of the spectra,
underpredict the PHENIX data 
by a factor of $\sim$2.  This is understood. As mentioned earlier,
Peterson's fragmentation functions are hard and it produces less number
of heavy mesons. 

\begin{figure}[h]
\begin{minipage}{18pc}
\includegraphics[bb=30 250 520 800,width=15pc,height=15pc]{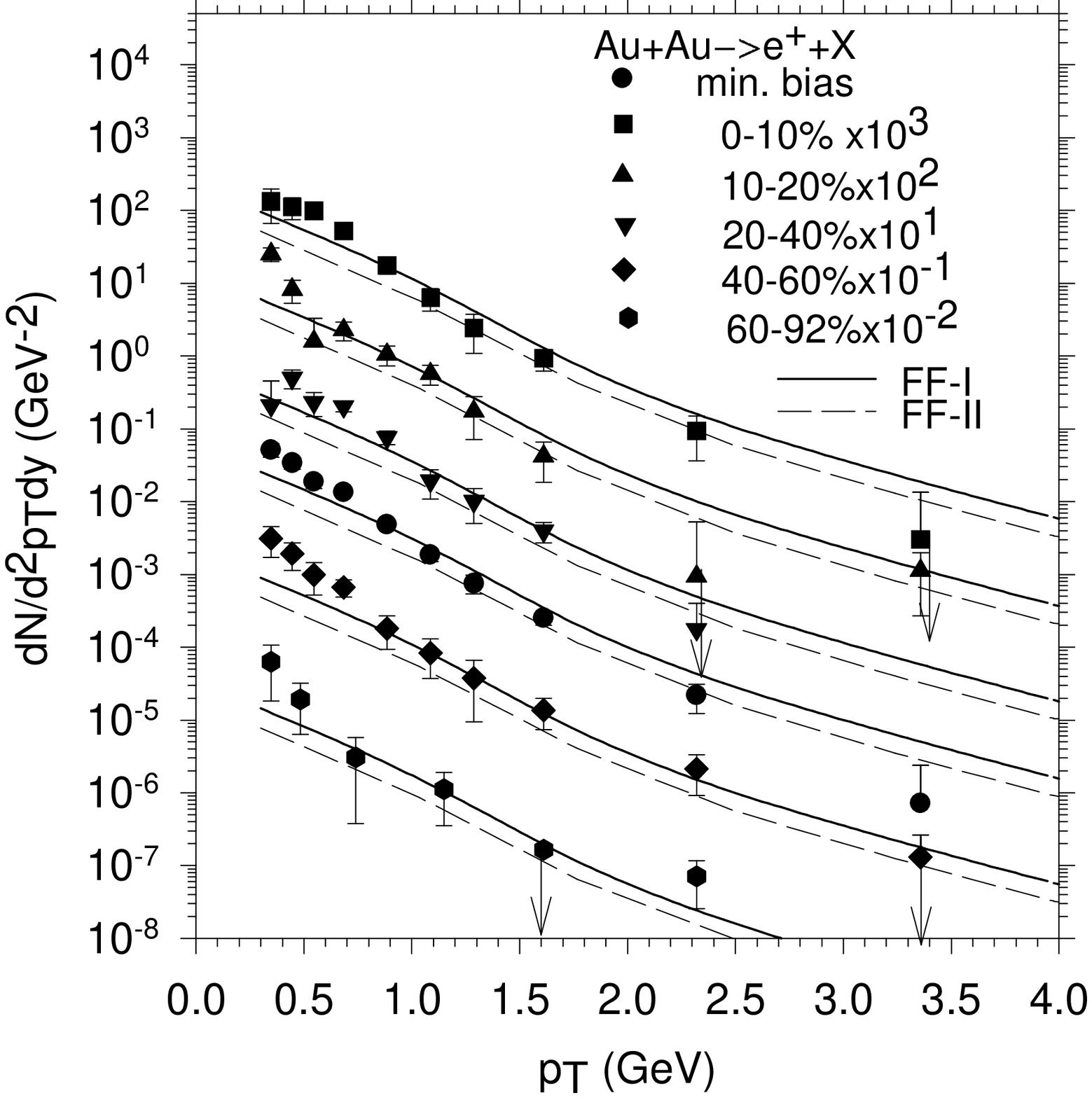}
\caption{\label{F7a}
Transverse momentum distribution of charmed
electrons in Au+Au collisions at RHIC energy, $\sqrt{s}$=200 GeV.
Solid lines are pQCD predictions (K=1.8) for p+p collisions, scaled by the nuclear
overlap function.}
\vspace{1cm}
\end{minipage}\hspace{2pc}%
\begin{minipage}{18pc}
\includegraphics[bb=30 250 520 800,width=15pc,height=15pc]{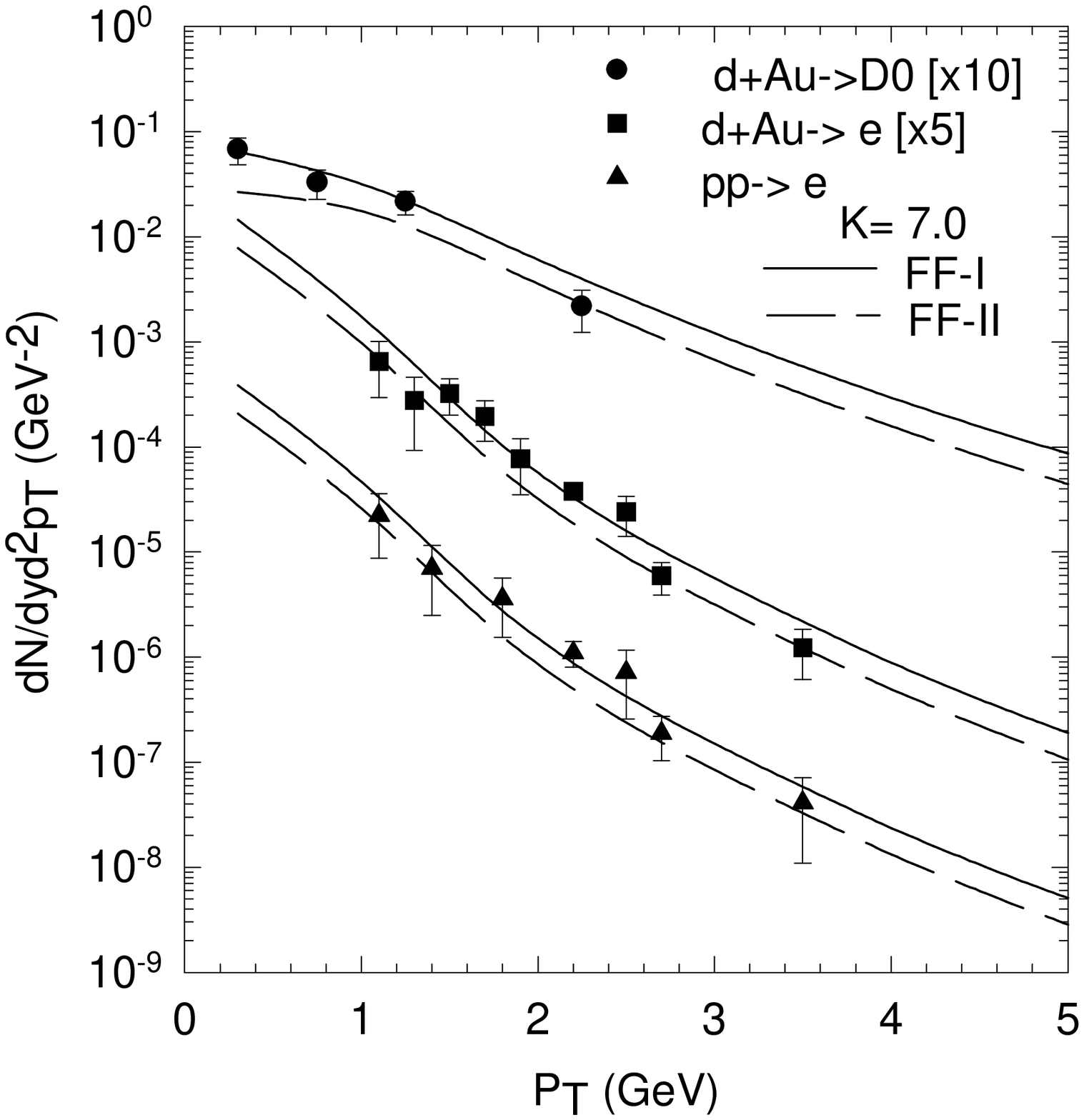}
\caption{\label{F8}
STAR measurements of 
$p_T$ distribution $D$-meson (solid circles) and charmed electrons
(solid squares) in d+Au collisions at RHIC energy $\sqrt{s}$=200 GeV.
STAR pp results for decay electrons are shown as solid triangle. The solid
and dashed lines are the pQCD predictions with FF-I and FF-II respectively
with K=7.0 .}
\vspace{8mm}
\end{minipage} 
\end{figure}

In Fig.3, PHENIX measurements \cite{phenix1} in Au+Au collisions, in different
centrality bins are shown. As mentioned earlier, charm electrons
scale with binary collision numbers. In Fig.3, present pQCD model
results, with  K=1.8 are shown. The solid and dashed lines are obtained with fragmentation function FF-I and FF-II respectively. We have included
the B meson decay contributions.
For comparing with PHENIX data the pp-predictions are multiplied
by the nuclear overlap function. As expected, model predictions
with Braaten parameterisation of fragmentation function  agree well
with the experiment. Here again, Peterson's parameterisation underpredict
the data approximately by a factor of 2. Analysis of PHENIX data 
confirms that  heavy quarks
suffer little or no energy loss in the medium produced in Au+Au collisions
at RHIC energy.

We now compare the present model predictions with the STAR experiment.
STAR collaboration \cite{star1} have measured $p_T$-distribution of $D$-mesons in d+Au collisions at
 $\sqrt{s}$=200 GeV. They have also
measured the semileptonic decay electrons in 
d+Au and p+p collisions.
In Fig.4, STAR data are shown. As mentioned earlier, for STAR experiment,
we use K=7.0 .
In Fig.4, solid and dashed lines present model predictions with fragmentation functions FF-I and FF-II respectively.  For  d+Au collisions, 
we
have  multiplied the results for pp predictions by $N_{bin}=7.5$.
We note that  the fragmentation function FF-I, with K=7 can describe
the STAR data on charm production in p+p, p+d collisions.
FF-II, as before leads to poorer description of the data.

\subsection{Transverse momentum distribution of $D^0$,$B^0$ and 
semileptonic decay electrons at LHC} 

We have shown that the present pQCD model with  the
heavy quark fragmentation function,
parameterised by Braaten et al \cite{br95}, 
 reproduces the STAR and PHENIX data on the transverse momentum
distribution of $D$ meson and semileptonic decay electrons
in $\sqrt{s}$=200 GeV p+p, d+Au and Au+Au collisions. 
With a reasonable
estimate of K, 
the model can be used to predict
heavy flavor production in pp collisions at LHC energy, 
As mentioned earlier, $K$ factor decreases with energy and 
at LHC energy, expected value is $K\sim 1-1.5$ \cite{es03}. 
In the present section, we give predictions for the transverse momentum distribution of D, B and semileptonic decay
electrons with K=1. We only show the predictions obtained with the fragmentation function FF-I, parameterised by Braaten et al \cite{br95}. As shown earlier, RHIC data are better explained with this parameterisation. Before we proceed, a note of caution is in order. The
model is tested against RHIC data, limited to $p_T \leq$ 4.5 GeV. 
Accuracy of the model beyond $p_T$=4.5 GeV is not tested. In the following, we will show model prediction over a wide range of $p_T$, but
the predictions for $p_T \geq$ 5 GeV may not 
be accurate.
   
\begin{figure}[h]
\begin{minipage}{18pc}
\includegraphics[bb=30 250 520 800,width=15pc,height=15pc]{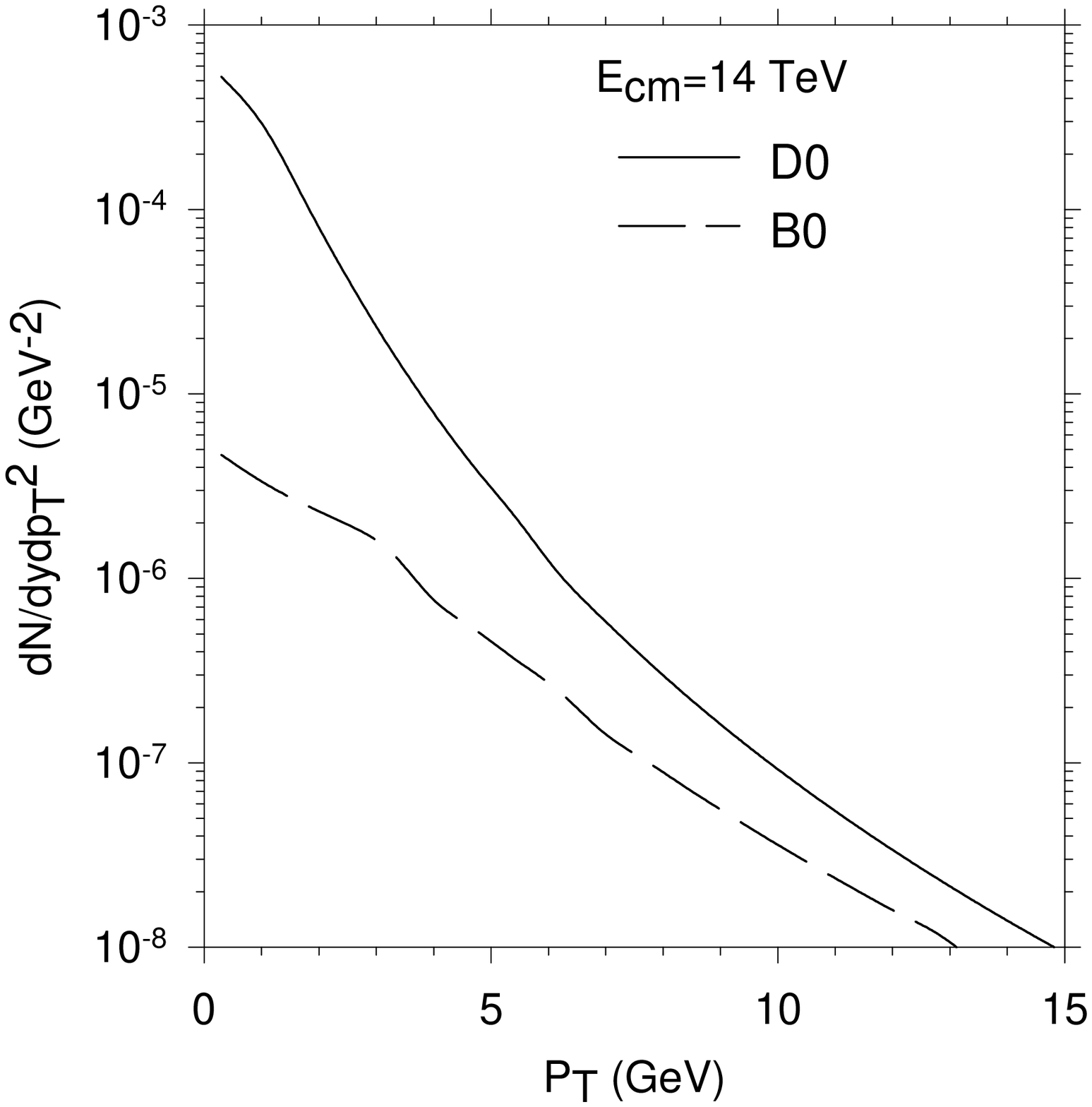}
\caption{\label{F9}
Transverse momentum distribution of
$D$ mesons (solid line) and $B$ mesons (dashed line) in pp collision
at LHC energy, $\sqrt{s}$=14 TeV.}
\vspace{1.5cm}
\end{minipage}\hspace{2pc}%
\begin{minipage}{18pc}
\includegraphics[bb=30 250 520 800,width=15pc,height=15pc]{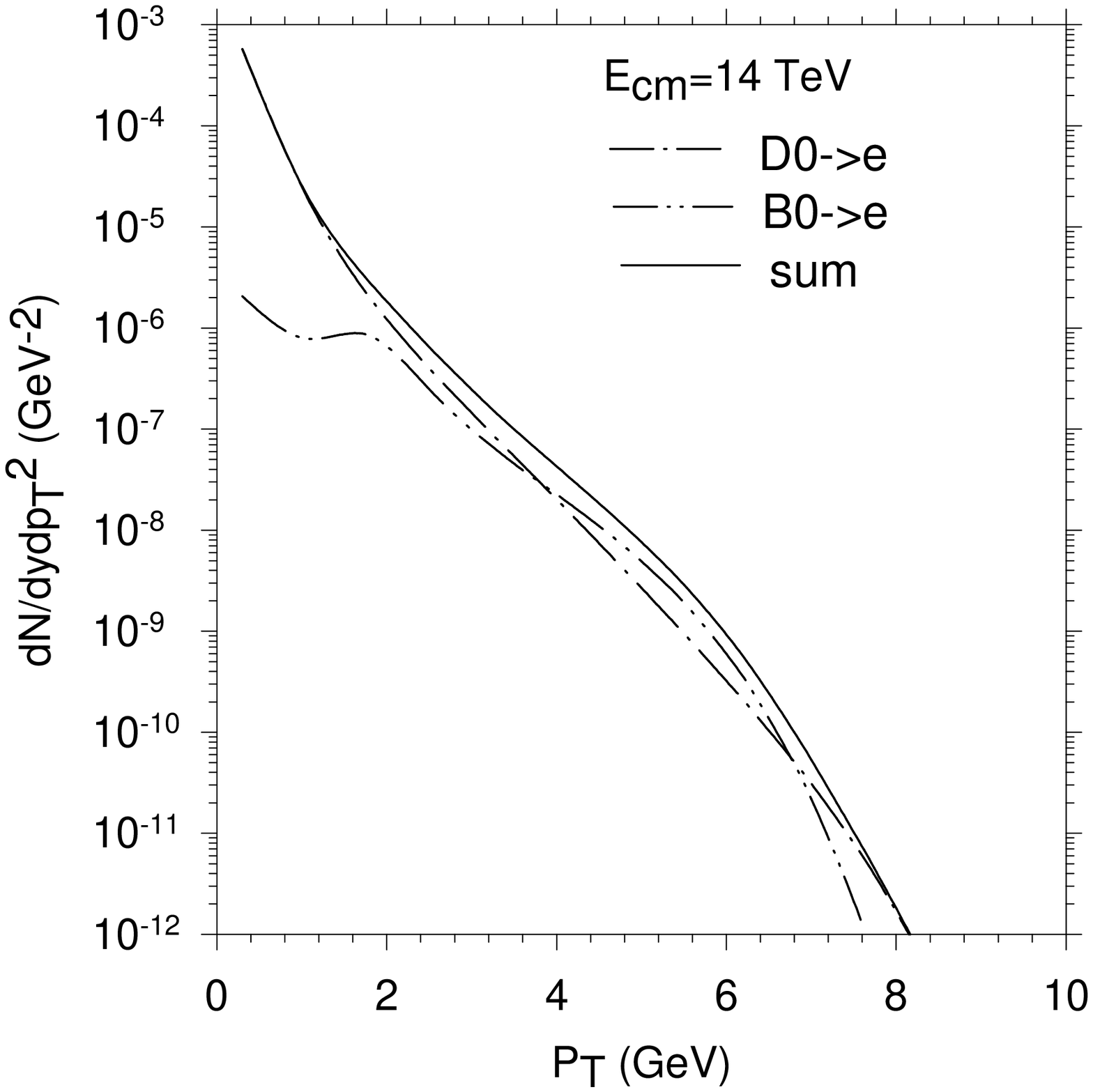}
\caption{\label{F10}
Transverse momentum distribution of electrons in semileptonic decay
of $D$ mesons (dash-dot line), $B$ mesons (dash-dot-dot line) in pp collisions
at LHC energy $\sqrt{s}$=14 TeV. The solid line is the sum of the two contributions.}
\vspace{8mm}
\end{minipage} 
\end{figure}

In Fig.5, model predictions,  for the transverse momentum distributions
of  $D$ and $B$ mesons 
in pp collisions at LHC energy $\sqrt{s}$=14 TeV, are shown.  
As it was at RHIC energy,
at LHC energy also, low $p_T$ heavy mesons are dominated by the
D-mesons. At low $p_T$ $B$-meson yield is significantly lower
than the D meson yield. Only at large $p_T$, B-meson contribute significantly to total heavy meson production.
However, compared to RHIC energy (see Fig.1), at LHC energy both $D$ and $B$ meson yields
are increased. High $p_T$ yields are significantly increased, by a factor,
1000 or more. 

In Fig.6, $p_T$ distribution of electrons from
semileptonic decay of D and B meson are shown. Low $p_T$ electron spectra 
is dominated by the electrons from decay of $D$-meson. Below
$p_T$=1 GeV, B mesons
contribute less than 10\% to the total yield. B meson contribution
increases rapidly with $p_T$ and in $p_T$ range 2-4 GeV, it rises to
30-50\%. However, at very large $p_T> 6 GeV$ B mesons contribution decreases again.

\section{summary and conclusions}
	 
In a pQCD based model, we have studied the heavy flavor production
at RHIC energy pp collisions. 
At RHIC energy, heavy flavor production
is dominated by the charm quarks. Below $p_T$=2 GeV, more than 90\%
heavy quarks are charm. Production of bottom quark picks up only at
large $p_T$ and at $p_T$=5 GeV, it contribute to 20\% of to total heavy quark yield.  As a consequence of dominance of charm quark
at low $p_T$, at RHIC low $p_T$ heavy mesons are dominantly D-mesons.
Similarly decay electrons are dominantly from decay of D meson.
We have compared the pQCD model predictions for $D$ and decay
electrons with STAR and PHENIX data. This models require a K-factor.
At RHIC energy, we fix the K-factor to reproduce the integrated charm yield 
$d\sigma^{c\bar{c}}/dy$, estimated in STAR and PHENIX experiments.
With K-factor fixed from total charm yield, the model, with fragmentation function parameterised
by Braaten et al \cite{br95}, reproduces the transverse momentum 
spectra of semileptonic decay electrons measured by the
PHENIX and STAR collaboration in $\sqrt{s}$=200 GeV p+p collisions.
pp predictions scaled by the nuclear overlap function also explain
the PHENIX Au+Au charm data in different centrality bins.
$p_T$ distribution of $D^0$ in p+d collisions, measured by the STAR collaboration is also explained in the model.
The results confirms that charm production in d+Au
and Au+Au collisions obey binary collision scaling law. The other,
much used fragmentation function parameterised by Peterson et al
\cite{pe83}, though reproduces the shape of the distributions, underpredict
the yields by a factor of $\sim$ 2-3.  
Assuming that at LHC energy, $\sqrt{s}$=14 TeV, K-factor approached unity,
we have used the model 
 to predict
heavy flavor production in p+p collisions at LHC energy. Compare to
RHIC energy,
at LHC energy collisions, heavy flavor production is increased.
At large $p_T$ increase is more than a factor 1000.  

\appendix
\section{pQCD model for $p_T$ distribution of semileptonic decay of $D\rightarrow Xe\nu$}

Production cross section of a heavy quark $c$ (charm)
in a pp collision can be written as \cite{st95}

\begin{eqnarray}  \label{a1}
\frac{d\sigma^c}{dyd^2p_T} = && K \sum_{i,j}^{partons} \int
dx_a  dx_b f_i(x_a,Q^2) f_j(x_b,Q^2) \frac{\hat{s}}{\pi}
 \frac{d\sigma }{d\hat{t}} (ab\rightarrow cd)
\delta(\hat{s}+\hat{t}+\hat{u}-m_a^2-m_b^2-m_c^2-m_d^2)
\end{eqnarray}
 
\noindent where $x_a$ and $x_b$ are the fractional momenta of the colliding partons.
$(\hat{s},\hat{t},\hat{u})$ are the Mandelstam variables for the subprocess
$ab \rightarrow cd$.
$\frac{d\sigma}{d \hat{t}}$ is the subprocess
cross-section. The factor $K$ takes into account the neglect of higher order
terms.

For a open charm production 
subprocesses $ab \rightarrow cd$ of interest are:

\begin{eqnarray}
gg \rightarrow C\bar{C} \label{p1}\\
q\bar{q} \rightarrow C\bar{C} \label{p2}\\
gQ \rightarrow gC\\
qQ \rightarrow qC
\end{eqnarray}

\noindent where $C$ denote a charm quark. g and q denote  a gluon and a light quark respectively.

Matrix elements for these processes are calculated in \cite{co79}.
For charm quark mass $M$ subprocess cross sections are,

\begin{mathletters}
\begin{eqnarray}
|M|^2_{q\bar{q} \rightarrow c\bar{c}} &&= \frac{64}{9} \pi^2 \alpha^2(Q^2) \frac{ (M^2-\hat{t})^2+(M^2-\hat{u})^2 + 2M^2\hat{s} }{\hat{s}^2 }\\
|M|^2_{gg \rightarrow c\bar{c}} &&= \pi^2 \alpha^2(Q^2) 
[\frac{12}{\hat{s}^2}(M^2-\hat{t})(M^2-\hat{u}) 
+ \frac{8}{3} \frac{(M^2-\hat{t})(M^2-\hat{u})-2M^2(M^2+\hat{t})}{(M^2-\hat{t})^2}\\ \nonumber
&&+\frac{8}{3} \frac{(M^2-\hat{t})(M^2-\hat{u})-2M^2(M^2+\hat{u})}{(M^2-\hat{u})^2} 
- \frac{2}{3} \frac{M^2(\hat{s}-4M^2)}{(M^2-\hat{t})(M^2-\hat{u})}\\ \nonumber
&& -6 \frac{(M^2-\hat{t})(M^2-\hat{u})+M^2(\hat{u}-\hat{t})}{s(M^2-\hat{t})} -6\frac{(M^2-\hat{t})(M^2-\hat{u})+M^2(\hat{t}-\hat{u}) }{\hat{s}(M^2-\hat{u})}]\\
|M|^2_{qc \rightarrow qc} &&=\frac{64}{9}\pi^2 \alpha^2(Q^2) \frac{(M^2-\hat{u})^2+(\hat{s}-M^2)^2+2M^2\hat{t}}{\hat{t}^2}\\
|M|^2_{gq \rightarrow gc} && = \pi^2 \alpha^2(Q^2) 
[ \frac{32(\hat{s}-M^2)(M^2-\hat{u})}{\hat{t}^2} 
+\frac{64}{9} \frac{(\hat{s}-M^2)(M^2-\hat{u})+2M^2(\hat{s}+M^2)}{(\hat{s}-M^2)^2}
\\
\nonumber
&&+\frac{64}{9} \frac{(\hat{s}-M^2)(M^2-\hat{u})+2M^2(M^2+\hat{u})}{(M^2-\hat{u})^2}
+ \frac{16}{9} \frac{M^2(4M^2-\hat{t})}{(\hat{s}-M^2)(M^2-\hat{u})}\\ \nonumber
&&+16\frac{(\hat{s}-M^2)(M^2-\hat{u})+M^2(\hat{s}-\hat{u})}{\hat{t}(\hat{s}-M^2)} -16\frac{(\hat{s}-M^2)(M^2-\hat{u})+M^2(\hat{s}-\hat{u}) }{\hat{t}(M^2-\hat{u})}]
\end{eqnarray}
\end{mathletters}

For the subprocesses $gg\rightarrow C\bar{C}$ and $q\bar{q}\rightarrow C\bar{C}$, we assume the following forms for the parton momenta  a,b and c.

\begin{mathletters}
\begin{eqnarray}
p_a =&&(\frac{x_a\sqrt{s}}{2}, 0, 0, \frac{x_a\sqrt{s}}{s})\\
p_b =&&(\frac{x_b\sqrt{s}}{2}, 0, 0, \frac{x_b\sqrt{s}}{s})\\
p_c=&&(M_T chy, p_T, 0, M_T sh y), 
M_T=\sqrt{M^2+p^2_T}
\end{eqnarray} 
\end{mathletters}
 
Mandelstam variables $\hat{s}=(p_a+p_b)^2$, $\hat{t}=(p_a-p_c)^2$ 
and $\hat{u}=(p_b-p_c)^2$, for these subprocesses
can be calculated as,

\begin{mathletters}
\begin{eqnarray}
\hat{s} = &&x_ax_b S \\
\hat{t}=&& M^2 - x_a \sqrt{s} M_T exp(-y) \\
\hat{u}=&&M^2 - x_b \sqrt{s}M_T exp(y)
\end{eqnarray}
\end{mathletters}

For the subprocesses involving heavy quark excitations, we assume the following
forms for $p_a$,$p_b$ and $p_c$

\begin{mathletters}
\begin{eqnarray}
p_a =&&(\frac{x_a\sqrt{s}}{2}, 0, 0, \frac{x_a\sqrt{s}}{s})\\
p_b=&&(\sqrt{M^2+\frac{x_b^2s}{2}}, 0, 0, - \frac{x_b\sqrt{s}}{2})\\
p_c=&&(M_T chy, p_T, 0, M_T sh y)
\end{eqnarray} 
\end{mathletters}

Corresponding Mandelstam variables are
 
\begin{mathletters}
\begin{eqnarray}
\hat{s} =&&M^2 + x_a x_b s\\
\hat{t}=&& M^2 - x_a \sqrt{s} M_T exp(-y) \\
\hat{u}=&&2M^2 -x_b\sqrt{s}M_T exp(y)
\end{eqnarray}
\end{mathletters}

One of the integration in Eq.\ref{a1} can be eliminated with the condition 
$\hat{s}+\hat{t}+\hat{u}=2M^2$, and charm production cross section can
be obtained as, 

\begin{eqnarray} \label{a2}  
\frac{d\sigma^C}{dyd^2p_T} = && \sum_{i,j}^{partons} \int^1_{x_a^{min}}
dx_a  f_i(x_a,Q^2) f_j(x_b,Q^2) \frac{2}{\pi}
\frac{x_ax_b+M_q^2/s}{2x_a-x_T exp(y)}
\frac{d\sigma }{d\hat{t}} (ab\rightarrow cd)
\end{eqnarray}
 
\noindent where
\begin{eqnarray}
x_b=&&\frac{x_a x_T e^{-y}- 4M^2_q/s}{2x_a-x_Te^{-y}}\\
x_a^{min}=&&\frac{x_T e^y - 4M^2_q/s }{2-x_Te^{-y}}
\end{eqnarray}

\noindent where $M_q=0$ for the subprocesses, $gg(q\bar{q}) \rightarrow C\bar{C}$ and $M_q=M_c$ for the subprocesses,
$gC\rightarrow gC$ and $qC\rightarrow qC$ .

 
If $D_{D/C}(z,\mu^2)$ is the fragmentation function for quark $C$ to fragment
into $D$ meson $h$,
transverse momentum distribution of $D$ are obtained as,

\begin{eqnarray}  
\frac{d\sigma^D}{dy_h d^2q_T} = &&\int \frac{dz}{z^2} D_{D/C}(z,\mu^2)
\frac{d\sigma^C}{dyd^2p_T}\\
=&& \sum_{i,j}^{partons}
\int_0^1 \frac{dz}{z^2} D_{D/C}(z,\mu^2)
\int_{x_a^{min}}^1
dx_a  f_i(x_a,Q^2) f_j(x_b,Q^2) \frac{2}{\pi}
\frac{x_ax_b+M_q^2/s}{2x_a-x_T e^y}
\frac{d\sigma }{d\hat{t}} (ab\rightarrow cd)
\end{eqnarray}
 
For a given hadron momentum ($q_T$) and rapidity ($y_h$), 
corresponding quantities ($p_T,y$) of 
the fragmenting parton are obtained from the relations,

\begin{mathletters}
\begin{eqnarray}
\sqrt{M_{h}^2 +q^2_T} ch y_h = &&z \sqrt{M^2+p_T^2} ch y\\
\sqrt{M_{h}^2 +q^2_T} sh y_h =  &&\sqrt{M^2+p_T^2} ch y
\end{eqnarray}
\end{mathletters}
 
Semileptonic decay of heavy mesons $D\rightarrow Xe\nu$ has been
studied in detail \cite{gr77,al79}. We will not elaborate on the procedure.
Transverse momentum distribution of
electrons from the decay of the D-mesons can be calculated as \cite{gr77}, 

\begin{equation}
\frac{dN^e}{dp_T}= \int \frac{dN^D}{dp^\prime_T} H(p_T,p^\prime_T) dp^\prime_T
\end{equation}

\noindent with

\begin{equation}
H(p_T,p^\prime_T) = \int \frac{d(p_T.p^\prime_T)}{2|p_T| p_T.p^\prime_T}
f(\frac{p_T.p^\prime_T}{M_D})
\end{equation}
 
\noindent where $f$ is the rest frame distribution of decay electrons. 
We have used the following rest frame spectrum for the decay 
$D\rightarrow Xe\nu$ \cite{gr77},

\begin{equation}
f(E_e)=\frac{1}{\Gamma}\frac{d\Gamma}{dE_e}= w g(E_e)
\end{equation}

\noindent where

\begin{eqnarray}
g(E_e)=&&\frac{E_e^2(M^2_D-M^2_X-2M_DE_e}{M_D-2E_e} \\
w=&&\frac{96}{(1-8m^2+8m^6-m^8-24m^4 \ln m)M_D^6}\\
m=&&M_X/M_D
\end{eqnarray}

We assumed that for D-mesons decay, $M_X=M_K$=0.497 GeV.

Above formulation is given in terms of charm quarks and D mesons,
They are also applicable for bottom quark and B-meson production
with appropriate change of masses. For
semileptonic decay of B-mesons, $B\rightarrow Xe\nu$, 
$M_X=M_D$. 

Decay electron spectra are appropriately weighted by the  
branching ratios. For $D\rightarrow e$
branching ratio is 0.103 and that for decay of $B\rightarrow e$
branching ratio is 0.109 \cite{ca05}.
We note that $B\rightarrow D \rightarrow e$ channel
also contribute to total electron spectrum, but we have ignored it,
as second order contribution.


\begin{references}  
\bibitem[*]{byline}e-mail:akc@veccal.ernet.in
\bibitem{star1}J. Adams et al [STAR Collab.], Phys. Rev. Lett. 94 (2005)062301 [arXiv:nucl-ex/0407006];
An Tai [STAR Collab.], J. Phys. G30 (2004)S809, [arXiv:nucl-ex/0404029].
\bibitem{phenix1}S. S. Alder et al [PHENIX collab.], Phys. Rev. Lett.94(2005)082301;
S. Kelly [PHENIX Collab.], [arXiv:nucl-ex/0403057].
\bibitem{star2} J. Adams et al, STAR Collaboration,
Phys.Rev.Lett.91, 2003,072304,nucl-ex/0306024.
Phys.Rev.Lett.91, 2003,172302,nucl-ex/0305015.
Phys.Rev.Lett.89, 2002,202301,nucl-ex/0206011.
\bibitem{phenix2}S. S. Alder et al, PHENIX collaboration,
Phys.Rev.C69,2004, 034910, nucl-ex/0308006.
Phys.Rev.Lett.91,2003,072303, nucl-ex/0306021.
Phys.Rev.Lett.91,2003,072301, nucl-ex/0304022
\bibitem{phobos1}B. B. Back et al., PHOBOS collab.
Phys.Rev.Lett.91,2003,072302, nucl-ex/0306025.
Phys.Lett.B578 2004, 297.
\bibitem{brahms1} I. Arsene, BRAHMS collab.,
Phys.Rev.Lett.91,2003, 072305.
\bibitem{gy90}M. Gyulassy and M. Plumer, Phys. Lett. B243, 1990, 432.
\bibitem{wa92}X. N. Wang and M. Gyulassy, Phys.Rev. Lett. 68 (1992) 1470.
\bibitem{ma86} T. Matsui and H. Satz, Phys. Lett. B178,416(1986).
\bibitem{th01}R. L. Thews, M. Schroedter and J. Rafelski, Phys. Rev. C63(2001)054905.
\bibitem{ca98}M. Cacciari, M. Greco and Paolo Nason, [arXiv:hep-ph/9803400].
\bibitem{vo02}R. Vogt, [arXiv;hep-ph/0203151].
\bibitem{ca05}M. cacciari, P. Nason and R. Vogt, [arXiv:hep-ph/0502203].
\bibitem{es03}K. J. Eskola and H. Honkanen, Nucl. Phys. A713, 167(2003).
\bibitem{br95}E. Braaten, K. Cheung, S. Fleming and T. C. Yuan,
Phys. Rev. D51 (1995)4819.
\bibitem{pe83}C. Peterson, D. Schlatter, I. Schmitt and P. M. Zerwas,
Phys. Rev. D27 (19982)105.
\bibitem{ka78}V. G. Kartvelishvili, A. K. Likhoded and V.A. Petrov,
Phys. Lett. B78(1978)615.
\bibitem{co85}P. Collins and T. Spiller, J. Phys. G11 (1985)1289.
\bibitem{sj01}T. Sjoestrand et al, Comput. Phys. Commun. 135(2001)238.
\bibitem{st95}G. Sterman et al, Rev. Mod. Phys, 67 (1995)157.
\bibitem{co79}B. L. Combridge, Nucl. Phys. B151(1979)429.
\bibitem{gr77}M. Gronau, C. H. Llewllyn Smith, T. F. Walsh, S. Wolfram and
T. C. Yang, Nucl. Phys. B123 (1977)47.
\bibitem{al79} A. Ali, Z. Phys. C1(1979)23, CERN-TH-2411

\end{references}
\end{document}